\documentclass[titlepage]{FLO_v1}%

\usepackage{graphicx}
\usepackage[authoryear]{natbib}
\usepackage{newtxtext}
\usepackage{newtxmath}

\usepackage[colorlinks,allcolors=blue]{hyperref}
\definecolor{jourcolor}{cmyk}{1,0.57,0.01,0.38}
\hypersetup{
    colorlinks,%
    citecolor=jourcolor,%
    filecolor=jourcolor,%
    linkcolor=jourcolor,%
    urlcolor=jourcolor
}

\articletype{RESEARCH ARTICLE}
\DOI{}
\Year{2023}
\Vol{}
\Price{}

\begin{document}

\title[]{Comparing flow-based and anatomy-based features \\ in the data-driven study of nasal pathologies}

\author[1]{Andrea Schillaci}
\author[1,2]{Kazuto Hasegawa}
\author[3]{Carlotta Pipolo}
\author[4]{Giacomo Boracchi}
\author[1]{Maurizio Quadrio$^\ast$}
\address[1]{Dept. of Aerospace Science and Technologies, Politecnico di Milano, Italy}
\address[2]{Department of Mechanical Engineering, Keio University, Japan}
\address[3]{Otolaryngology Unit, Asst Santi Paolo e Carlo, Department of Health Sciences, University of Milan, Italy}
\address[4]{DEIB, Politecnico di Milano, Italy}
\corres{*}{Corresponding author. E-mail: \emaillink{maurizio.quadrio@polimi.it}}


\date{\textbf{Received:} XXX; \textbf{Revised:} XXX; \textbf{Accepted:} XXX}

\abstract{
In several problems involving fluid flows, Computational Fluid Dynamics (CFD) provides detailed quantitative information, and often allows the designer to successfully optimize the system, by minimizing a cost function. 
Sometimes, however, one cannot improve the system with CFD alone, because a suitable cost function is not readily available: one notable example is diagnosis in medicine. The field of interest considered here is rhinology: a correct air flow is key for the functioning of the human nose, yet the notion of a functionally normal nose is not available, and a cost function cannot be written. 
An alternative and attractive pathway to diagnosis and surgery planning is offered by data-driven methods.
In this work, we consider the machine-learning study of nasal pathologies caused by anatomic malformations, with the aim of understanding whether fluid dynamic features, available after a CFD analysis, are more effective than purely geometric features in the training of a neural network for regression.
Our experiments are carried out on an extremely simplified anatomic model and a correspondingly simple CFD approach; nevertheless, they demonstrate that flow-based features perform better than geometry-based ones, and allow the training of a neural network with fewer inputs, a crucial advantage in fields like medicine.
}

\maketitle

\begin{boxtext}
\textbf{\mathversion{bold}Impact Statement}
Machine-learning (ML) algorithms and Computational Fluid Dynamics (CFD) techniques are often discussed together in the recent scientific literature in fluid mechanics. However, ML is always used as a tool to perform a better/cheaper/faster CFD. In this work, we explore the potential of the inverse approach, in which CFD provides useful information to a ML model. In an idealized three-dimensional problem, flow features restricted to the boundary of the computational domain and derived from CFD are shown to be more informative than the geometry of the boundary itself, leading to a better ML classifier, which can be trained with fewer labeled data. 

The application described in the paper is of the medical type, and concerns rhinology, where large amounts of accurately labeled data are not always available. Since the larger information content of flow-based features derives from the non-linear relationship between geometry and the corresponding flow field, the present result is relevant to other flow problems addressed with CFD where the lack of a clearly defined cost function suggests a data-driven approach.
\end{boxtext}

\section{Introduction}
\label{sec:intro}

With the continuous development of computing hardware and software, Computational Fluid Dynamics (CFD) is becoming increasingly useful in several applications, extending from industry to health. 
CFD, ranging from the cheaper and lower-fidelity flow models like the Reynold-averaged Navier--Stokes equations (RANS) to the opposite extreme of the highly accurate direct numerical simulation, is a useful tool to improve the design of industrial systems, by e.g. increasing the aerodynamic efficiency of an airplane, reducing the aerodynamic drag of a vehicle, or enhancing the mixing in a fluidic system. 

Sometimes, however, the CFD solution, albeit informative, does not explicitly provide an immediate means to improve the system.
This is often the case in the medical field. The specific example considered in this work concerns the air flow in the human nose. 
The nasal cavities are the connecting element between the external ambient and the lungs, and serve a number of additional functions, which include smell, filtering and humidifying the incoming air, and heating/cooling it to the correct temperature. Most of these functions are directly driven by the anatomical shape of the nasal cavities. 
In fact, nowadays Ear, Nose and Throat (ENT) surgeons routinely take their surgical decisions mostly based on the analysis of Computed Tomography (CT) scans, which provide a detailed view of the anatomy of the nasal cavities.
In principle, the nose flow can be well described by CFD, which is indeed increasingly used to support ENT doctors in their diagnosis \citep{moreddu-etal-2019, tjahjono-etal-2022}, and to improve our understanding of the complex physics of the nose flow \citep{calmet-etal-2019, farnoud-etal-2020}. 
Yet, the basic questions routinely asked by the ENT doctors (e.g. whether to perform a surgery on a given patient, and where) cannot straightforwardly answered by CFD alone. Designing a surgery can be considered to be akin to a shape optimization problem. Unfortunately, the mathematical and numerical tools available for shape optimization cannot be deployed to solve the nose problem, because the goal is not self-evident, owing to the lack of a functionally normal reference nose. A huge anatomical variability among healthy anatomies is present, which makes the discrimination between healthy and pathological cases far from obvious.

Several studies have attempted to develop a robust workflow for a CFD analysis of the nose flow \citep{quadrio-etal-2014, tretiakow-etal-2020}, and to understand what is a healthy airflow \citep{zhao-jiang-2014, borojeni-etal-2020} via multi-patient analyses. However, as seen from a clinical perspective, the present state of affairs remains unsatisfactory. There is evidence that the rate of failure of certain surgical corrections is extremely high: for the correction of septal deviations, for example, more than 50\% of the patients report poor postoperative satisfaction ratings \citep{rhee-etal-2003, sundh-sunnergren-2015, tsang-etal-2018}. Although there is general agreement \citep{inthavong-etal-2019} that CFD offers a significant potential for improved surgery planning, this potential remains as yet largely untapped.

An interesting approach to diagnose pathologies and suggest surgeries relies on the use of Machine Learning (ML) techniques. The central question that we are going to address in this paper is whether CFD-computed flow information can be useful in this process, and possibly be more effective than the geometrical information embedded in the CT scan. 

ML in fluid mechanics has recently seen a huge activity and recorded significant progresses \citep[see e.g. the review by][]{vinuesa-brunton-2022}; however, very little information is available in the literature regarding the combined use of ML and CFD in rhinology, if exception is made for our own preliminary study \citep{schillaci-etal-2021}. The use of artificial intelligence and machine learning techniques has been limited so far to classification of images derived from CT scans \citep{crowson-etal-2020}.
In closing their paper, \cite{lin-hsieh-hsieh-2020} mention that putting together CFD and ML would be an interesting future avenue for research. 
A very recent study by \cite{jin-etal-2023} uses CFD and ML approaches, but only one at a time, and there is no attempt to combine them in any way. 

The main goal of the present work is to answer the question whether flow features derived from CFD of the nasal airflow can be useful for a ML-based analysis. 
The possibility that CFD-based features outperform geometrical ones is rooted in the nature of the highly non-linear Navier--Stokes equations, which link anatomy and the ensuing CFD solution.
A positive answer would carry general interest in all those situations where ML is used in the context of flow systems to replace optimization, since a cost function is not readily available. 

To answer the question above, we consider synthetic (but realistic) healthy and pathological nasal anatomies, created with CAD, and a simple CFD solution of the flow within them (computed with the Reynolds-averaged Navier--Stokes equations and a standard turbulence model). An inference model made by a standard neural network is trained to understand whether each synthetic nasal anatomy is affected by a pathology, and to predict its severity. Two alternative approaches are employed. One resembles the approach currently followed by ENT doctors for their diagnosis, and is solely based upon geometric/anatomic information. The other, instead, relies on flow features extracted from CFD. 

By comparing the performance of the two types of features, we intend to understand whether CFD, albeit somewhat costly, carries potential advantages, like e.g. increased accuracy or the need for less observations, over anatomy alone. In particular, the ability to train the ML model with information derived from fewer patients would be extremely important, owing to the difficulty of  obtaining highly informative and accurately labeled training data in health-related applications.

The structure of the work is as follows. After this Introduction, \S\ref{sec:method} describes the generation of a suitable set of geometries, the setup of CFD simulations, the extraction of information from the CFD solution, and the neural network used for regression. Results presented in \S\ref{sec:experiments} are used to critically discuss the comparison between geometrical and CFD features. Lastly, concluding remarks are put forward in \S\ref{sec:conclusions}.

\section{Methods}
\label{sec:method}

This Section illustrates the entire workflow, and describes in \S\ref{sec:anatomy} how the parametric CAD geometries of the noses are created, in \S\ref{sec:CFDsetup} how the CFD simulations are set up, in \S\ref{sec:FM} how different anatomies and flow solutions are compared, and in \S\ref{sec:NN} how a neural network is designed and trained to predict nasal pathologies.

A CAD-based, synthetic reference nose model is built first. The model presents the fundamental advantage of being parametric; changing the numerical values of a small set of geometric parameters allows us to introduce anatomical variability, related to both physiological inter-subject differences and pathological conditions. 
Using this parametric nose model, 200 distinct anatomic shapes are generated. For each shape, a point-to-point mapping is computed between each nose and the reference nose model.

A CFD simulation is then carried out for each geometry. Thanks to the previously computed mapping, both flow and geometrical features of each geometry can be mapped back to the reference one, so that any (flow or geometrical) feature of any model can be observed on the reference one. 
At this point, a neural network is trained to perform a regression on the geometrical parameters of each nose and to predict the value of the pathological parameters.

\subsection{The anatomies}
\label{sec:anatomy}

All the anatomies considered in the present work originate from a reference CAD-based simplified model of the human nasal cavities, already introduced by \cite{schillaci-etal-2021}, which lends itself to a simple geometrical parametrization. 
The use of simplified model geometries for the study of the flow in the human nasal cavities is not new: for example \cite{liu-etal-2009} employed a model obtained by averaging together the CT scans of 30 patients. Our CAD-based approach relates more closely to that by \cite{naftali-etal-1998}, who used a CAD nose-like model to reproduce the essential features of the nasal cavities. However, we are first to build a fully parametrized model, that is required for the generation of a complete and controlled dataset.

\begin{figure}
\centering
\includegraphics[width=0.8\textwidth]{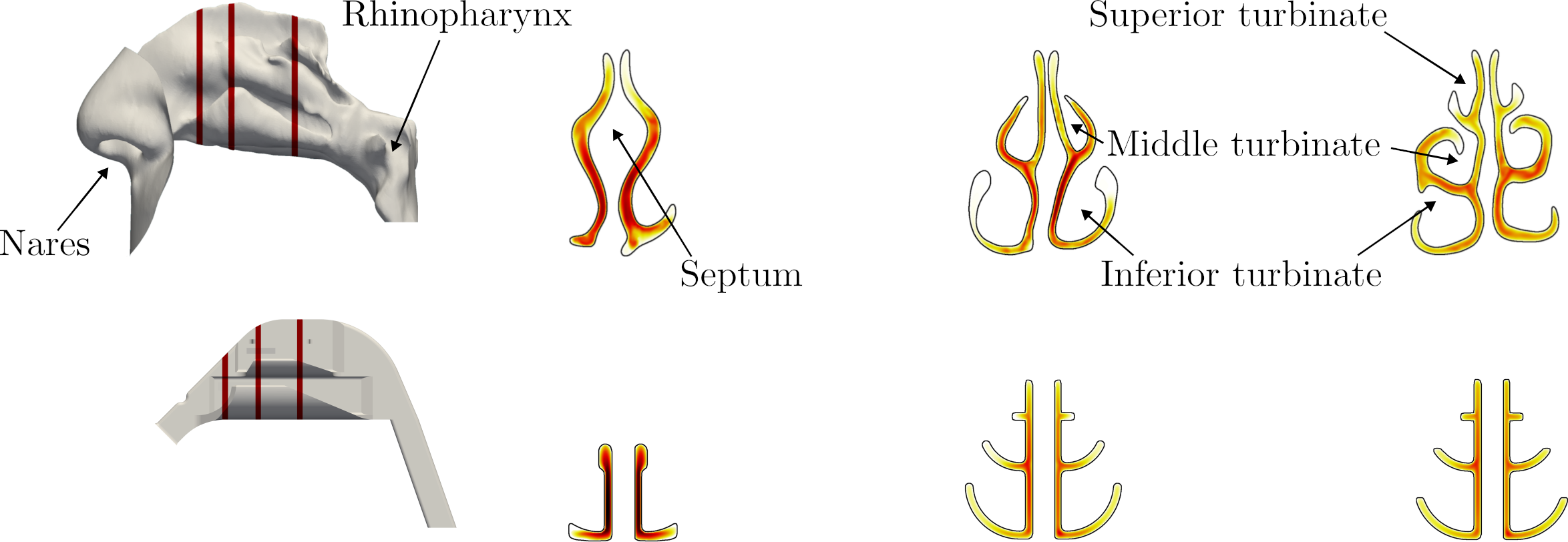}
\caption{Comparison between a real nasal anatomy (top) and the simplified CAD model used in the present work (bottom). The three-dimensional view on the left marks the three coronal sections plotted on the right. The cross-sections with the colormap represent the CFD solution in terms of the magnitude of the velocity vector. The top solution results from averaging a time-averaged LES solution (not reported in this paper), and the bottom one is a RANS solution. Although the CAD model is highly simplified, the major features are in line with the real anatomy.}
\label{fig:nose_anatomy}
\end{figure}

The reference CAD nose model is meant to represent an healthy anatomy, and realistically mimics the major anatomical structures of a real nose. A qualitative comparison between the nasal cavities of a real patient (top) and our simplified baseline model (bottom) can be seen in figure \ref{fig:nose_anatomy}. The CAD model, developed in collaboration with a group of ENT surgeons, is designed to reconcile the opposite requirements of simplicity and clinical significance. Its planar or constant-curvature surfaces are indeed highly idealized, but the model replicates in a quantitatively accurate way the crucial anatomical features, such as the dimensions of the septum between the two fossae, the hook-like structure of the inferior and middle turbinates, and the thickness of the passageway in the most critical areas of the nasal fossae. This is an essential prerequisite to provide deformations with clinical significance.
The comparison (shown in figure \ref{fig:nose_anatomy}) of the flow field computed with an high-fidelity approach on a patient-specific anatomy confirms the suitability of the present model.

The nasal geometry begins anteriorly with the nares, and ends posteriorly with the rhinopharynx and the hypopharynx. 
The nasal septum, a thin structure lying approximately on the median plane, separates the nasal cavities in two halves, the left and right nasal fossae. Each fossa has a cross-sectional shape that changes significantly along the nasal vestibules, developing as a narrow channel of convoluted shape, medially bounded by the septum; the particular conformation of the fossae is considered to improve humidification and heat exchange. 
Three long and curled bony structures, the (inferior, middle and superior) turbinates, define the cross-sectional shape of the fossae. The turbinates unfold roughly parallel to the flow, and are attached to the lateral walls of the nose. The inferior turbinate is the largest, running almost the entire way from the vestibule to the rhinopharynx. 

\begin{figure}
\centering
\includegraphics[width=0.7\textwidth]{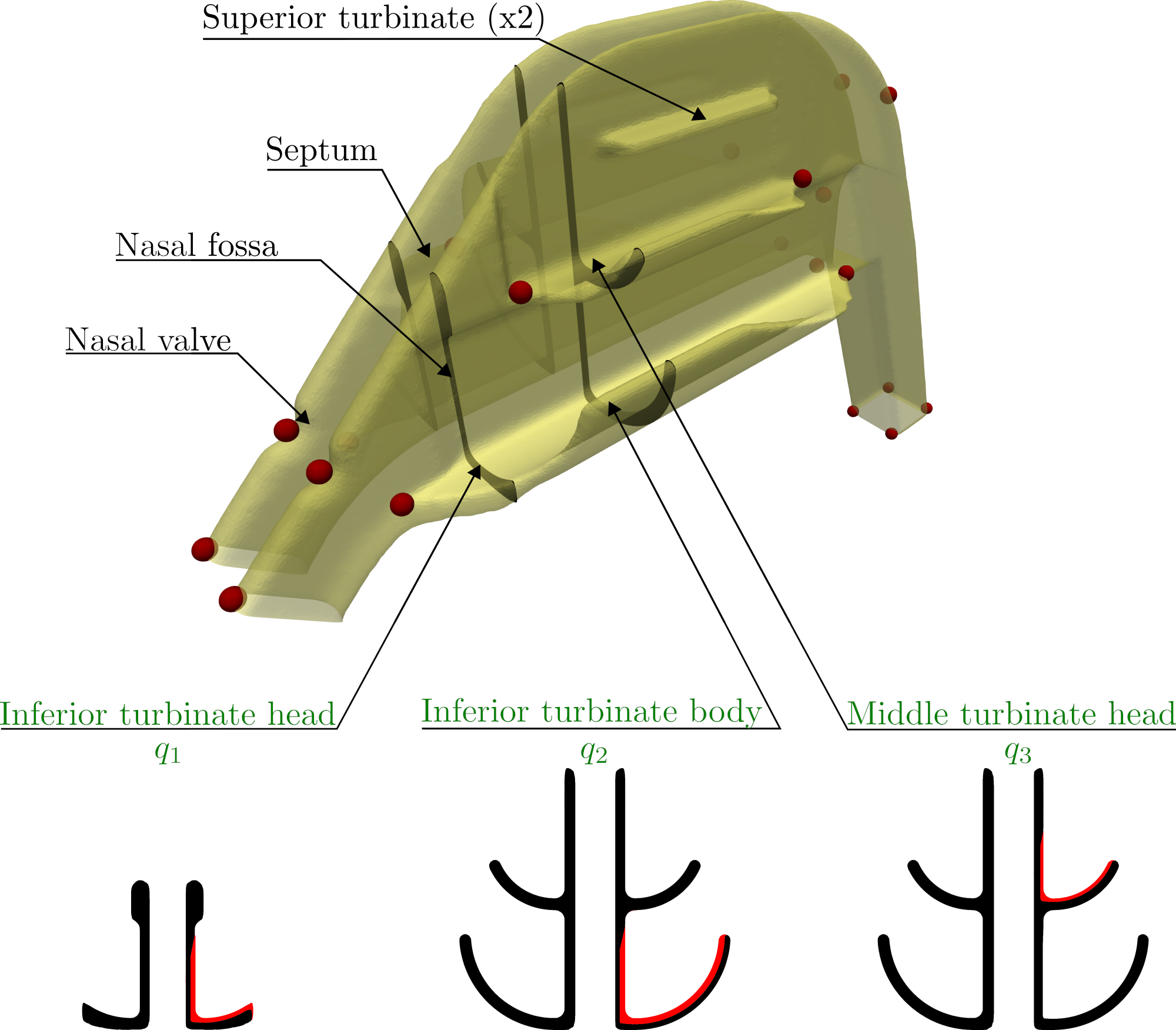}
\caption{Nose model and position of its parametric modifications. Top: three-dimensional view, with black text labels indicating physiological variations, and green text highlighting pathologies. The red points are landmarks used for functional mapping. Two cross-sections (corresponding to the two most anterior ones plotted in figure \ref{fig:nose_anatomy}) are plotted on the bottom, and highlight in red the geometry changes induced by the pathological parameters $q_1$, $q_2$, and $q_3$.}
\label{fig:pathologies}
\end{figure}

The parametric CAD model contains eight numerical parameters, whose range of variation is meant to account at the same time for the physiological and the pathological variability of real anatomies. Three parameters $q_1$, $q_2$ and $q_3$ correspond to three clinically sensitive regions of the nasal cavities, and describe the intensity of selected pathologies, related to an hypertrophy of the turbinate; the remaining five parameters are clinically harmless, and represent the physiological variability among healthy anatomies. The parameters describe  anatomical changes or their arbitrary combinations; their numerical values are expressed in millimeters and are changed in steps of 0.05 $mm$.
In the healthy reference anatomy, all the parameters are set to zero. Figure \ref{fig:pathologies} shows where in the model the parameters act to modify the reference nose. 

Ninety-nine extra healthy anatomies are created from the reference one by varying the values of the five healthy parameters. They alter (see figure \ref{fig:pathologies}) the vertical and longitudinal position of the superior turbinate, the axial position of the nasal valve, the thickness of the septum in the area of the nasal valve, and the thickness of the septum in correspondence to the head of the inferior turbinate. Varying the parameters produces localized and relatively small changes in the geometry: the strongest change leads to a 14\% reduction in cross-sectional area in the most affected section, while the smallest change decreases the area by 0.7\% only. 

One hundred pathological anatomies are created by varying the values of the remaining three pathology-related parameters. They are designed to mimic in a quantitatively reliable way a common condition called turbinate hypertrophy, a swelling of the turbinates which produces a constriction of the meati, up to a point where the airway may, in the most severe cases, become completely obstructed. The obvious consequence is reduced nasal patency and difficulty to breathe.
Such hypertrophies have a number of causes, from allergic rhinitis to inflammation of the sinuses, and affect one or more turbinates in either of the nasal fossae. 
The reference CAD shape is modified by altering the values of the three patological parameters; their numerical values are set under tight supervision of ENT doctors, and remain within clinically meaningful values, to ensure that deformations are realistic, notwithstanding the idealized model.
Parameter $q_1$ varies between 0 and 0.7 and mimics an hypertrophy of the head (anterior portion) of the inferior turbinate; parameter $q_2$ varies between 0 and 0.7 and  mimics an hypertrophy of the body (intermediate portion) of the inferior turbinate; parameter $q_3$ varies between 0 and 0.55 and mimics an hypertrophy of the head of the middle turbinate. 
All the pathologies are applied to the right fossa. 
The geometrical changes induced by non-zero pathological parameters are visualized in the bottom part of figure \ref{fig:pathologies}.

An important point to stress is that the parameterization of the geometry provides an unambiguous label for each case. The label consists in numerical values of the three pathological parameters: in other words, for each case it is precisely known which pathology is at play, and how much is its severity.
This characteristic, that will be essential when training the inference model, is impossible to obtain when working with real anatomies.

\subsection{Simulations}
\label{sec:CFDsetup}

A CFD solution is computed for each of the 200 distinct anatomies. 
In view of the simplified geometries employed here, a basic RANS flow model is employed: the finite-volume package OpenFOAM \citep{weller-etal-1998}, with its SIMPLE solver and a first-order discretization are used to arrive quickly at a converged solution. In doing so, we follow standard modeling and discretization choices, which are summarized below.

The computational grid is generated with the utilities available in OpenFOAM. An uniform background mesh with cubic cells of side length $1 \ mm$ is created first; the mesh is then refined further and adapted to the surface. The final mesh is rather coarse, in line with the RANS approach, and consists of around 1.1 millions cells. Values in the nasal flow literature range between 0.1 up to 44 millions cells \citep{inthavong-etal-2018}, with the finest meshes being generally used in highly resolved LES studies \citep{calmet-etal-2016, covello-etal-2018}, meanwhile typical, modern RANS simulations range between 1 and 4 millions elements \citep{liu-etal-2007, wen-etal-2008}. Although its significance in the present problm is limited, we mention that the average value of $y^+$ is $0.6$, and maximum and minimum values are $4e-3$ and $1.9$. No wall treatment is used, neither in terms of boundary conditions (i.e. no wall functions) nor damping functions. The mesh adopts no cell layers near the wall, and relies on the decrease of cell size induced by the snappyhexmesh tool in presence of irregular boundaries. 

The dataset is built for a steady inspiration, which is considered as the most clinically representative breathing condition. The inspiration is driven by a pressure difference of $20 \ m^2/s^2$ between the inlet and the outlet section, which for the reference geometry corresponds to a flow rate of about  $178 \ ml/s$. This value corresponds to an inspiration at rest or at mild physical activity \citep{wang-lee-gordon-2012}. 
As in the vast majority of CFD studies in this field \citep{radulesco-etal-2019}, the nasal walls are considered as rigid, thus neglecting the compliancy of the tissues (which is very small at this low breathing rate) and the modifications of the erectile mucosal tissue during the nasal cycle, known to cyclically alter the shape of the passageways over a time scale of few hours \citep{patel-etal-2015}. 
The velocity has zero gradient at the inlet and outlet sections. A no-slip boundary condition for the velocity components is applied at the walls, whereas for pressure a zero-gradient condition is enforced. 
The turbulence model of choice is $k - \omega - SST$ \citep{menter-kuntz-langtry-2003}, which is commonly employed in such simulations \citep{li-etal-2017}. The model solves two additional differential equations, one for the turbulent kinetic energy $k$ and one for the turbulent frequency $\omega$. At the inlet $k$ is set by assuming just 1\% turbulent intensity, zero gradient is used at the outlet, and $k=0$ is set at the walls.  
For the turbulence frequency $\omega$, at the inlet $\omega=1 s^{-1}$ is prescribed, at the outlet a null gradient is imposed, and at the wall the value is set as in \cite{menter-1994}. 

The outcome of a typical simulation for the healthy anatomy is compared in figure \ref{fig:nose_anatomy} with the temporally averaged solution obtained on a real anatomy with an higher-fidelity (and significantly more expensive) CFD method, namely Large Eddy Simulation (LES), where a WALE turbulence model is used \citep{ducros-etal-1999}, for the same flow conditions. The LES mesh is of about 12.8 millions cells. Although the comparison clearly has to be intended in a qualitative sense only, it is seen that most of the flow passes through the same regions, in particular in the meatus between the septum and the inferior and middle turbinate for all the section. This confirms the suitability of the simplified model for the purpose of the present work.

We stress once again that, in the present study, seeking the highest fidelity in the solution is not our primary concern. Hence, the solution method (the RANS equations), the numerical schemes (first order) and the quality of the mesh (fairly coarse) are all standard, and meant to generate quickly and cheaply a dataset of reasonable size. 
Since each case carries around 100 $MB$ of data, the full dataset has a total size of 20 GB. To put these numbers into perspective, the Google Open Images Dataset V6 dataset \citep{kuznetsova-etal-2020} consists of around $9 \times 10^6$ images and is made by about 561 GB of data (including labels). In other words, our dataset is only one order of magnitude smaller, but consists of four order of magnitudes less observations, which emphasizes the high dimensionality of a typical CFD dataset.

\subsection{Functional maps}
\label{sec:FM}

A correspondence needs to be determined between the reference nose and each of the other noses in the full set. This correspondence is computed with a tool derived from computational geometry and called functional maps (FM) \citep{ovsjanikov-etal-2012}; functional mapping is briefly introduced below, and then specialized to the implementation employed here \citep{melzi-etal-2019}. 

Functional mapping provides an efficient method to estimate the correspondence between two shapes, as well as the correspondence of functions represented on them. It is a relatively new tool, that has been recently introduced for solving shape classification problems \citep{magnet-etal-2023}. 
Rather than directly estimating point-to-point correspondence between shapes, FM registers functional spaces defined over the two shapes. The multi-scale basis for the function space on each shape is given by the finite (truncated) set of eigenfunctions $\Phi_j, j=1 \ldots N$ of its Laplace--Beltrami operator. Once the basis is known, any function $f$ defined on the shape is approximated by the following linear combination of eigenfunctions
\[
f = \sum_{j=1}^N \gamma_j \Phi_j .
\]

In general, the functional mapping between two shapes is described by a matrix $A$, whose elements describe how each eigenfunction on one shape is expressed as a linear combination of the eigenfunctions on the other shape. We refer the interested readers to the original paper by \cite{ovsjanikov-etal-2012}, or to the recent contribution by \cite{magnet-ovsjanikov-2023}, for detailed information on functional mapping.

Given two shapes, each functional map is computed by solving a least squares minimization problem. To retrieve more precise maps, in our case twenty landmarks are identified for each of the shapes and used as descriptors. Landmarks, shown in figure \ref{fig:pathologies} for the reference nose, are points selected on the geometry because of their anatomical significance; they are often used in the ENT practice \citep[see e.g.][]{denour-etal-2020} to help dealing with different anatomies in the context of CT analysis or registration. 
As for the basis, the eigenfunctions of the Laplace--Beltrami operator used here possess the convenient characteristic of bringing out the dominating "frequencies" of the shape; therefore, they naturally provide a multi-scale representation of the geometry. 
Note that each nasal geometry has its own basis, but the more two geometries are similar, the more their bases are similar. Hence, the matrix $A$ becomes more diagonally dominant when two geometries are alike.

The specific FM implementation used here is called zoom-out, and has been introduced by \cite{melzi-etal-2019}. Instead of computing the map for the full set of basis functions, with zoom-out one computes first a smaller map and a smaller matrix that involves fewer basis elements (say, 10); the map is then iteratively extended, by adding rows and columns to $A$, up to the desired size. We have determined that, with the shapes of interest, truncating the modal expansion to $N=150$ eigenfunctions provides satisfactory results. Since the mapping is always between the reference shape and every other shape, 199 maps in total are computed.

Computing the correspondence between a generic shape and the reference one using FM involves the following steps:
\begin{enumerate}
    \item Compute (once) the truncated Laplace--Beltrami base on the reference shape;
    \item Compute the truncated Laplace--Beltrami base on every other shape;
    \item Position the 20 landmarks on each shape;
    \item Compute the functional map matrix $A$ by solving a least-squares optimization;
    \item Convert the matrix $A$ into a point-to-point map.  
\end{enumerate}

\begin{figure}
\centering
\includegraphics[width=\textwidth]{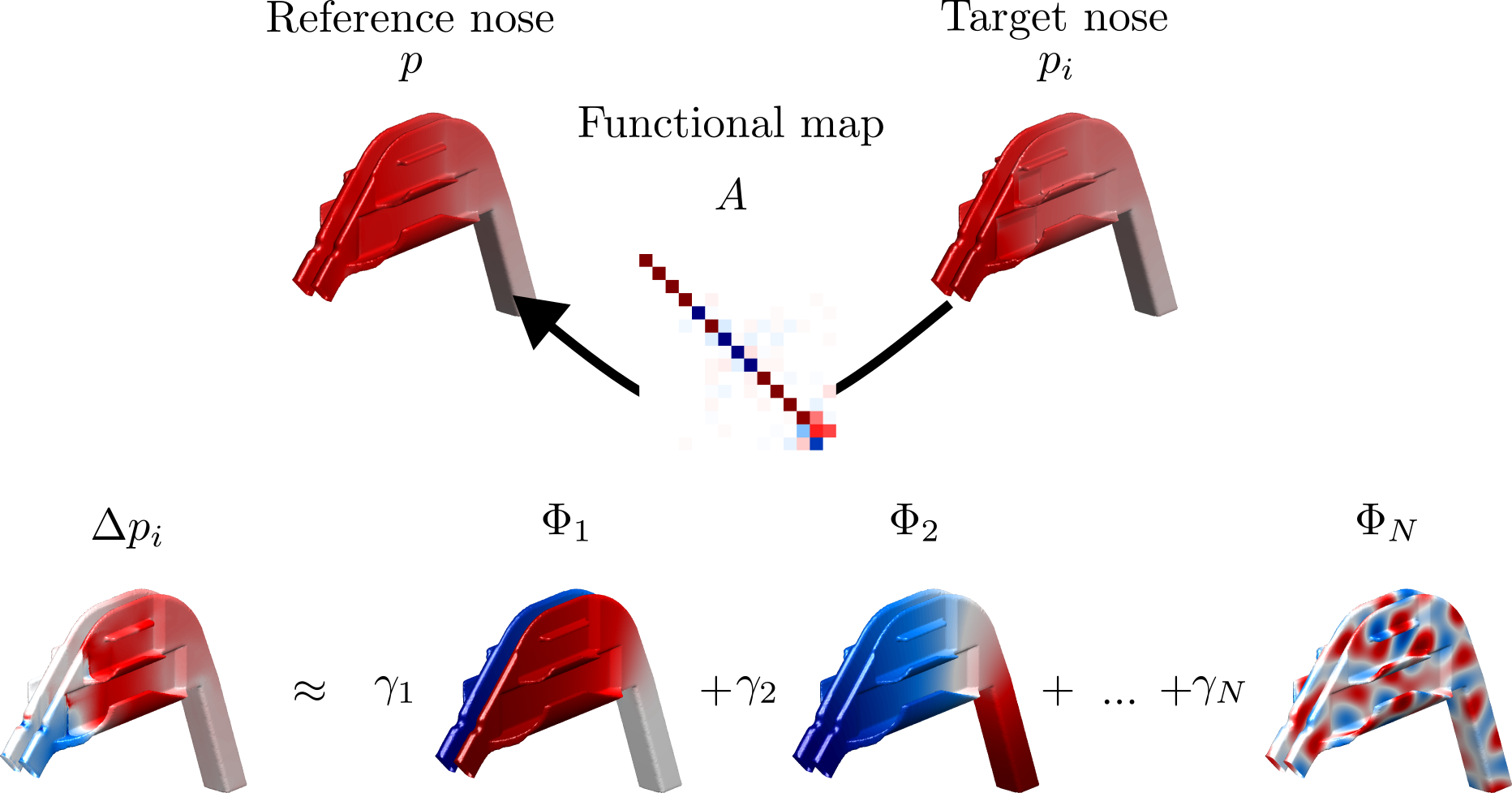}
\caption{Difference of a wall-based flow quantity (pressure $p$ in this figure) between the reference and the generic $i$-th anatomies. Pressure $p_i$ on the boundary of the $i$-th nose is mapped to the baseline nose as $\hat{p}_i$. The difference $\Delta p_i = p - \hat{p}_i$ is expressed as a linear combination of the eigenfunctions of the reference nose with coefficients $\gamma_j, j=1 \ldots N$.}
\label{fig:workflow}
\end{figure}

Once the matrix $A$ is available for the generic $i$-th nose, the workflow, graphically sketched in figure \ref{fig:workflow}, starts from the corresponding CFD solution, from which relevant flow quantities (for example, the pressure field $p_i$, or the skin-friction field $\tau_i$) are computed at the wall. Thanks to FM, the wall field $p_i$ is transported back to the reference nose to yield the field $\hat{p}_i$; the difference field $\Delta p_i = p - \hat{p}_i$ can be expanded by using the Laplace--Beltrami basis $\Phi$ of the reference anatomy and the corresponding coefficients $\gamma_j$, that will be used later in the regression. Note that eigenfunctions of the reference nose only are involved in the latter expansion, and that only wall-based quantities are mapped, thus eliminating any issue regarding the continuity equation. Vector field are transformed component-wise.

\subsection{The classifier}
\label{sec:NN}

Given the dataset with $\ell=200$ observations, a vector of input features (be it geometrical or derived from the flow solution) must be associated to a target value which describes the pathology through the numerical values of each the pathology parameters $q$. Carrying out the proper association is the task of a regressor, usually implemented as some kind of neural network (NN). The raw data from CFD for each observation are available in each cell of the discretized domain. 
Since the cardinality of raw data is much larger than $\ell$, a process of dimensionality reduction of the input, called feature extraction, is necessary to balance the number of observations with the number of inputs. 
The outcome of the feature extraction process depends on the specific experiment; hence, feature extraction will be described later in \S\ref{sec:experiments}, where the various experiments are discussed.

We train two different NN models, depending on the input data: a Multi-Layer Perceptron (MLP) \citep{goodfellow-etal-2016} when the input is a feature vector, and a Convolutional Neural Network (CNN) \citep{lecun-etal-1998} when the input is the matrix $A$, that we treat as spatial data. 
CNN is chosen because of its ability to capture patterns in maps; it is widely used in the field of image recognition, and it has also been applied to fluid dynamics in recent years \citep[see e.g.][]{fukami-fukagata-taira-2020}, due to its capability to deal with spatially coherent information. 

In general, designing the architecture of a NN involves several choices, e.g. deciding on the number of hidden layers and nodes, the activation function and the loss function. 
Our MLP is a regression network; it has an input layer, whose number of nodes is equal to the length of the feature vector, three hidden layers (with 30, 20 and 10 nodes each) and an output layer with one node only. The activation function is the hyperbolic tangent for all the nodes, except for the output nodes, which has a linear activation function. Since the goal is to predict the numerical values of the parameters which quantify the severity of the pathology, we adopt the mean square error as loss function. Lastly, the optimization algorithm, which updates weights and biases of the NN, is the classic Levenberg--Marquardt \citep{lera-pinzolas-2002}.
Reduction in the number of inputs is obtained by extracting the 20 most informative features via the Lasso method \citep{tibshirani-1996}.


Our CNN has a rather standard architecture.
The functional map, i.e. matrix $A$, is passed into a convolutional block, which consists of a 3-by-3 convolution layer, then a dropout layer randomly deactivates 20\% of the weights in its layer. 
Afterwards the data is normalized by a batch normalization layer and then fed into a hyperbolic tangent layer; finally data are reduced in size through a maxpooling layer with a 2-by-2 filter. The dropout and batch normalization layers are applied to avoid overfitting. 
After the convolution block, data are flattened into a vector and input into a fully-connected (FC) block, consisting of 20 perceptrons, dropout layers, batch normalization and hyperbolic tangent layers. The size of the FC block is halved until a single output node remains. The weights of the CNN are optimized with the widely used Adam method \citep{kingma-ba-2017}, owing to its efficiency and stability.

For both NN architectures, a reliable assessment of the error over the entire dataset is obtained with the $k$-fold cross-validation method \citep{james-etal-2021}. The dataset is partitioned over $k=5$ folds: each has 140 cases used for training, 20 for validation and 40 for testing. 
The 40 simulations used for testing do not overlap over the 5 folds, so that the performance is assessed over the whole dataset, albeit by training 5 different NN (with the same architecture). To avoid the potential bias of considering just a specific run, this operation is run 100 times per feature and per pathology, and eventually the average absolute error is computed.

\section{Experiments}
\label{sec:experiments}

Results of the regression experiments are now presented in comparative form, for geometrical and flow features. The goal of the experiments consists in retrieving the numerical value of the three pathological parameters.

In consideration of the relatively small size of the database, instead of training a single NN to predict the three parameters in a single attempt, we opt for training one NN for each of the three parameters.

\subsection{Geometrical features}

Multiple options exist to select geometrical features. In this work, we consider two features, identified with G1 and G2. The geometry-based feature G1 is simply the displacement between each point on a certain nose model and the corresponding point on the reference shape. Feature G2, instead, is the matrix $A$ obtained via FM when registering each nose with the reference one. 
Owing to the different nature of G1 and G2, a different NN architecture is used, namely a CNN for G2 and a MLP for G1. 

\begin{figure}
\centering
\includegraphics[width=0.8\textwidth]{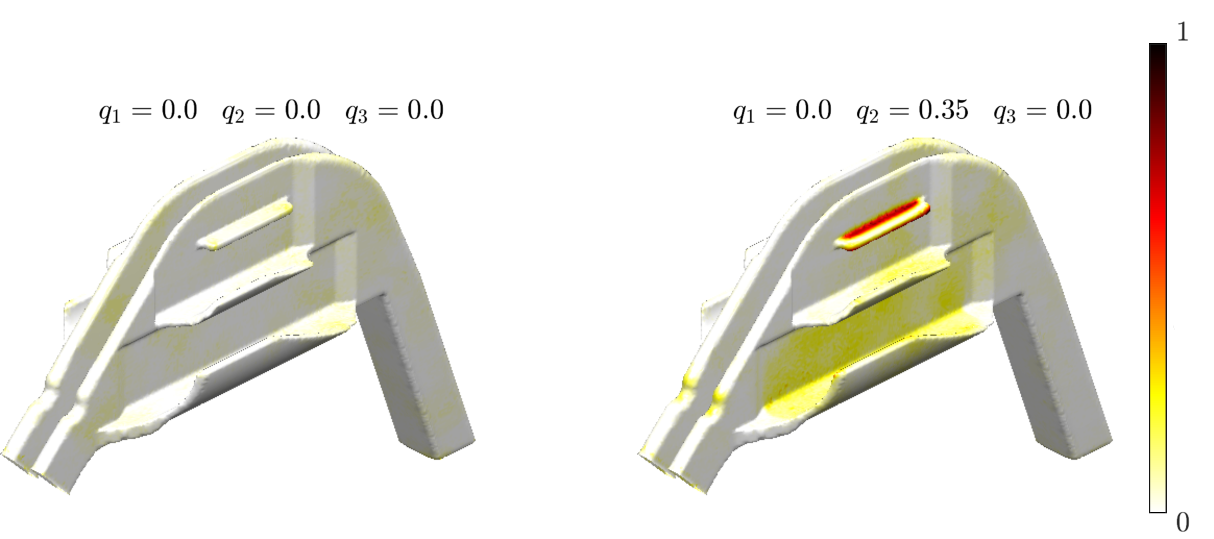}
\caption{Pointwise distance (feature G1) for an healthy nose (left) and a pathological nose (right). The numerical values of the three pathological parameters are reported above each panel. The right plot shows large values of G1 which highlight a non-pathological parameter (namely the position of the superior turbinate), and non-zero but smaller values in the area interested by the pathological parameter $q_2$ (which mimics hypertrophy of the body of the inferior turbinate). The colormap units are $mm$.}
\label{fig:G1}
\end{figure}

Feature G1 (shown in figure \ref{fig:G1}) is simply the distance between points on the reference nose and the corresponding ones on a modified shape. Since the modified shape is only varied through local changes, and thus does not need rgistration, unchanged points lead to zero displacement. 
The $i$-th nose is mapped on the reference nose, and the pointwise distance is a scalar field defined on the surface on the reference shape. 
This field is then decomposed into the Laplace--Beltrami basis ${\Phi_j}$ of the reference nose, arranged in a rectangular matrix, and the vector of Laplace--Beltrami coefficients ${\gamma}_j$. The ensuing overdetermined linear system is solved by using the Lasso method with a penalization constant. A careful choice of the constant leads to a solution with 20 coefficients only.

G1 is effective at spotting differences introduced by changes in the parameters. Figure \ref{fig:G1} indeed shows that G1 peaks exactly where the geometrical variations have been introduced. A quantity like G1 is expected to be subject to a small amount of noise, since each nose undergoes its own meshing process. Even though a certain portion the surface is unaltered, its points would modify their position slightly when a different mesh is computed. Nevertheless, figure \ref{fig:G1} shows that the noise remains more than acceptable, so that the deterministic geometrical changes are clearly highlighted: G1 peaks at the superior turbinate (determined by a non-zero value of a non-pathological parameter), and is also large in correspondence of the inferior meatus, defined by the inferior turbinate, which is affected by a pathology since $q_2=0.35$.

\begin{figure}
\centering
\includegraphics[width=\textwidth]{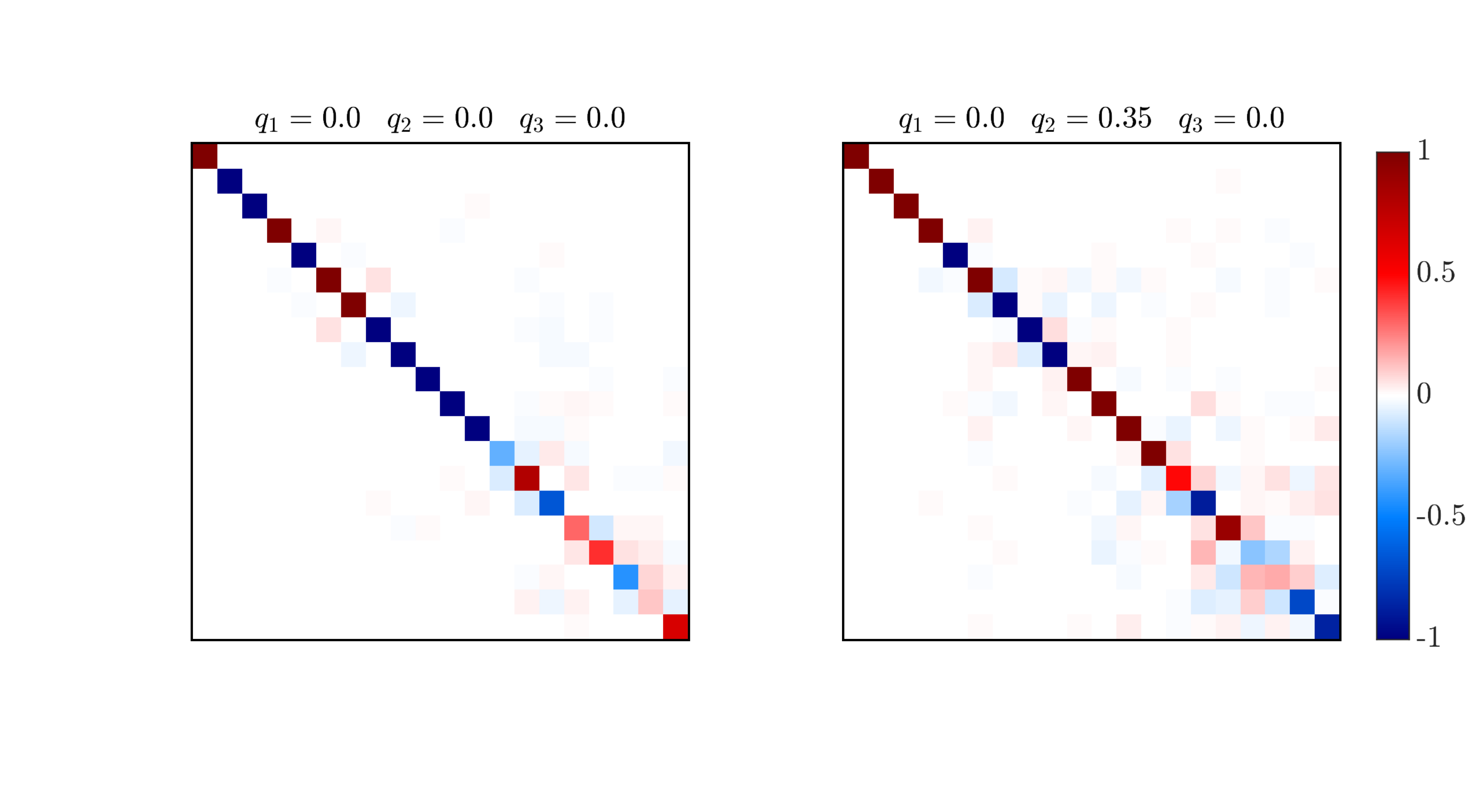}
\caption{Functional map (feature G2), represented by the sub-matrix $A_1$, for an healthy nose (left) and a pathological nose (right); each element is color-coded after normalization at unitary maximum. The numerical values of the three pathological parameters are reported above each panel. Healthy anatomies tend to produce more diagonal maps, whereas pathologies alter the diagonal structure of the matrix.}
\label{fig:G2}
\end{figure}

Feature G2, instead, involves the functional map between the two shapes, expressed via its matrix $A$. Owing to the usual need to avoid overfitting, the CNN is not given the full matrix $A$, but only a square sub-matrix $A_1$ of size 20. Hence, we are only comparing the first 20 eigenfunctions between the two anatomies. 
We stress again that $A$, since it is based on geometry only, does not contain information concerning the flow field, and is computed without the need of a CFD solution. 

Matrix $A$ appears to have a more diagonal structure when the shape corresponds to an healthy nose, compared to pathological cases. 
This can be confirmed by looking at figure \ref{fig:G2}, which portraits in graphical form the structure of the sub-matrix $A_1$: each square represents an element, whose value is encoded in its color. A diagonal matrix implies that each eigenfunction in one geometry is fully described by the corresponding eigenfunction in the other geometry. When off-diagonal elements are non-zero, one eigenfunction on one nose becomes a linear combination of several eigenfunctions on the other.

\subsection{Flow features}

\begin{figure}
\centering
\includegraphics[width=\textwidth]{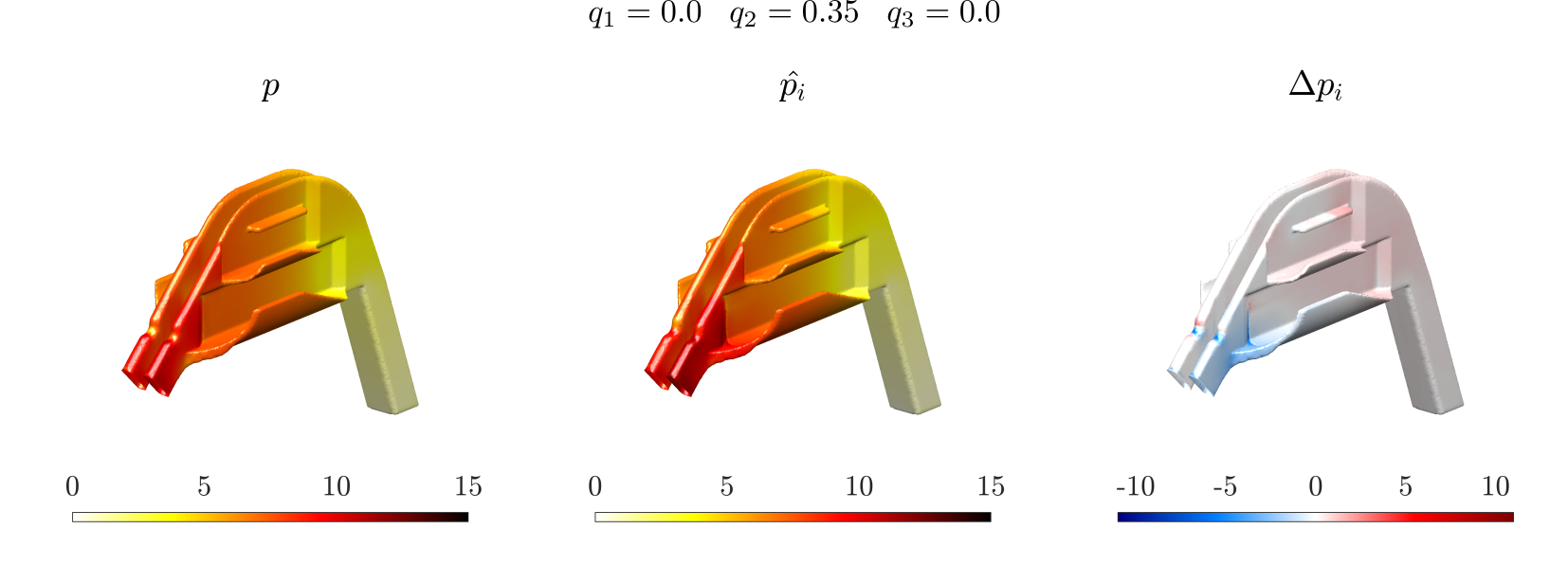}
\caption{Wall pressure $p$ (feature F1). Left: $p$ on the reference anatomy. Centre: $\hat{p}_i$ from a pathological anatomy (hypertrophy of the inferior turbinate) mapped back on the baseline. Right: $\Delta p_i$.  The numerical values of the three pathological parameters are reported above each panel. Colormap units are $Pa$.}
\label{fig:F1}
\end{figure}

\begin{figure}
\centering
\includegraphics[width=\textwidth]{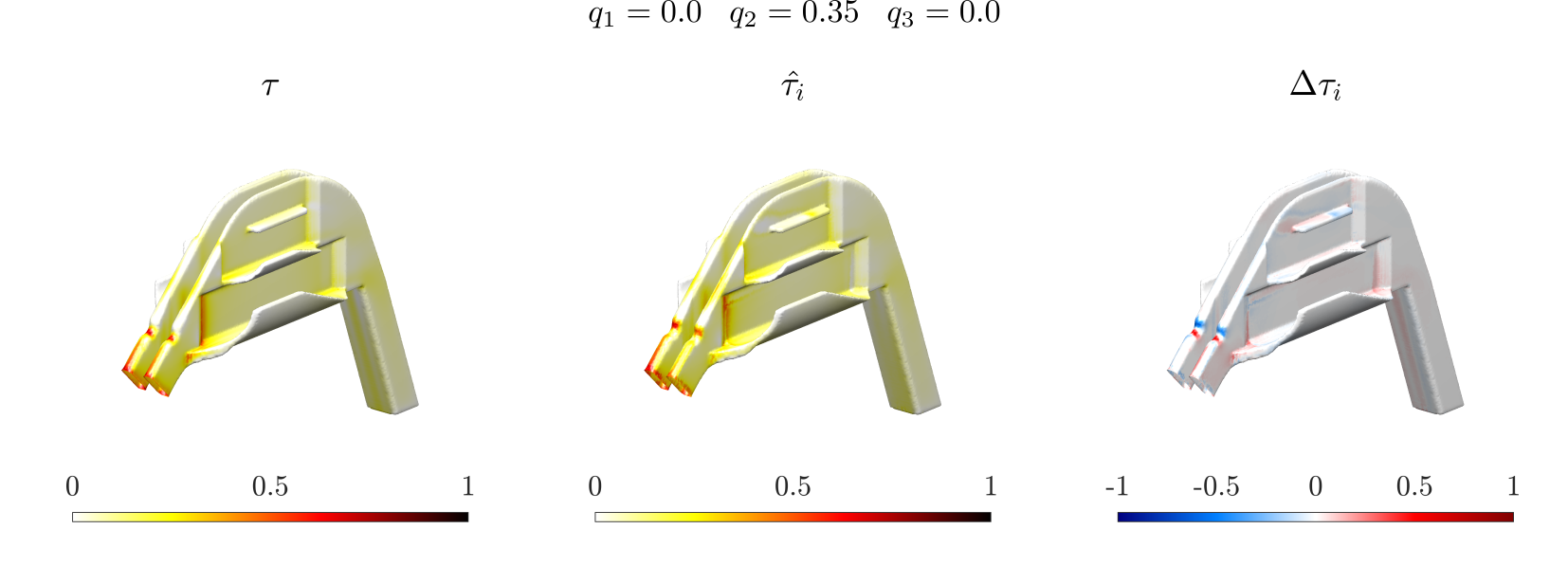}
\caption{Wall shear stress $\tau$ (feature F2). Left: $\tau$ on the reference anatomy. Centre: $\hat{\tau}_i$ from a pathological anatomy (hypertrophy of the inferior turbinate) mapped back on the baseline. Right: $\Delta \tau_i$. The numerical values of the three pathological parameters are reported above each panel. Colormap units are $Pa$.}
\label{fig:F2}
\end{figure}

Flow features are restricted at the wall; this choice is supported by clinical considerations: the feeling of discomfort is conveyed by  nerve terminations residing in the mucosal lining \citep{sozansky-houser-2014}. 
The most straightforward wall-based features one can resort to in an incompressible flow are wall pressure $p$ and the magnitude $\tau$ of the wall shear stress \citep{bewley-protas-2002}. As before, the difference of either quantity between the reference case and each mapped-back case is computed and expended in the reference basis: input to the NN are the first 20 coefficients of the expansion. 
In sharp contrast to geometry-based features, that have no need for CFD, these features are based on CFD but do not retain direct information about the geometry: only the eigenfunctions of the Laplace--Beltrami basis on the reference nose are used in the procedure. 

Figure \ref{fig:F1} illustrates the first flow-based feature F1, given by wall-pressure: it shows the wall-pressure field $p$ for the baseline nose (left), the transformed pressure field $\hat{p}_i$ for the $i$-th nose, affected by a severe hypertrophy of the head of the inferior turbinate (center), and the corresponding difference $\Delta p_i$ represented on the baseline nose. Although the two pressure fields $p$ and $p_i$ appear very similar, the difference field shows values of about $5 \ Pa$, which is a clinically significant value whose order of magnitude captures the variance in nasal resistance for patients with hypertrophy. 

Analogously, figure \ref{fig:F2} describes the second flow-based feature F2 for the magnitude of the wall shear stress $\tau$.

\begin{figure}
\centering
\includegraphics[width=\textwidth]{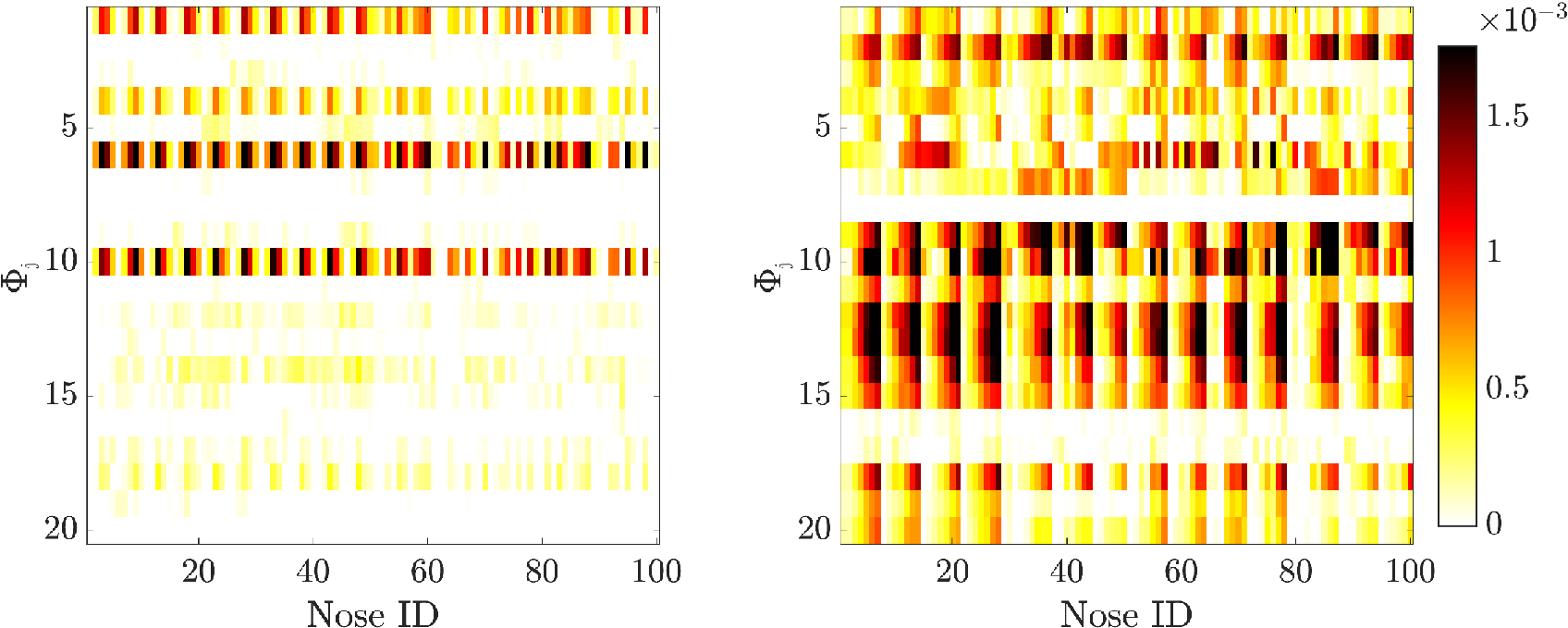}
\caption{Coefficients of the first 20 Laplace--Beltrami eigenfunctions of the expansion for the wall friction, for the 100 healthy noses (left) and the 100 pathological ones (right). Nose ID is on the horizontal axis, and mode number on the vertical axis (starting from the top).}
\label{fig:wssDetail}
\end{figure}

It is instructive to observe how wall-based flow quantities are reconstructed using the Lapace--Beltrami basis. 
As an example, for the wall shear stress feature F2, figure \ref{fig:wssDetail} compares the 100 healthy noses (left) with the 100 pathological ones (right): for each nose, the coefficients of the first 20 eigenfunctions of the Lapace--Beltrami expansion of the baseline anatomy are plotted, with color indicating the magnitude of the coefficient. 
The informative nature of the flow-based coefficients is apparent. It can be noticed that the pathological cases involve contributions from a higher number of modes, when compared to the healthy cases. Furthermore, certain eigenfunctions (e.g. modes 2,3,7,9) are minimally involved in the description of the flow solution for healthy noses, but become important to reconstruct the fields pertaining to pathological cases.


\subsection{Performance and discussion}

\begin{table}
\centering
\begin{tabular}{lcccc}
  Feature           & NN type                    & $q_1$ & $q_2$ & $q_3$ \\ 
\hline
G1: distance          & \multicolumn{1}{c|}{MLP} & 0.156   & 0.116  & 0.094  \\
G2: matrix $A_1$      & \multicolumn{1}{c|}{CNN} & 0.148   & 0.117  & 0.099  \\    
F1: pressure          & \multicolumn{1}{c|}{MLP} & 0.032   & 0.023  & 0.062  \\
F2: wall shear stress & \multicolumn{1}{c|}{MLP} & 0.019   & 0.019  & 0.041  \\ 
\hline
\end{tabular}
\caption{Average value (in millimeters) of the test error computed over 100 runs, for each of the three pathological parameters $q_1$, $q_2$ and $q_3$. The test error of one run contains the average results of the 5 NN derived from $k$-fold cross-validation with $k=5$. Flow-based features show a consistently lower error than geometry-based features.}
\label{tab:results}
\end{table}

Results of the regression experiments are presented in table \ref{tab:results}, in terms of the average error over the whole dataset. Distinct experiments are carried out for each pathology. 
The first, striking and most important observation is that geometry-based features produce an error which is about one order of magnitude larger than that obtained with flow-based features. 
Within flow-based features, the wall shear stress achieves a better performance over pressure in a consistent way over the set of pathologies. The pathology parameter $q_3$ (hypertrophy of the middle turbinate) turns out to be the hardest to predict via flow features (although its error when geometric features are used is the lowest among the three pathologies): this is reasonable, because an hypertrophy of the middle turbinate affects the anatomy in a region that is not crucial in terms of distribution of the flow (see figure \ref{fig:nose_anatomy}).  
However, even for such an unfavorable situation, flow features do perform significantly better the geometric ones.

The reason for the superiority of flow-based features can be traced back to the pathology being a geometrical modification which is typically quite localized in space, at least in our simplified model. As such, pathologies are small-scale features that tend to be visible only as higher-order modes of the Laplace--Beltrami basis. 
However, higher-order modes might be {\em per se} quite noisy, and dependent on the geometrical discretization. This difficulty could be further emphasized by the pathologies considered here, which aim at being clinically faithful and as a consequence sometimes involve geometric modifications which are small in absolute terms, often a fraction of a millimeter. 
If pathologies are examined in terms of the corresponding flow field, instead, these small changes are passed through the "filter" of the Navier--Stokes equations: a small and localized geometrical change in a sensitive position leads to a larger, more easily identified modification of the flow field. 
This, in turn, tends to appear within lower-order modes of the Laplace--Beltrami basis, and thus becomes easier to capture.

\begin{figure}
\centering
\includegraphics[width=0.75\textwidth]{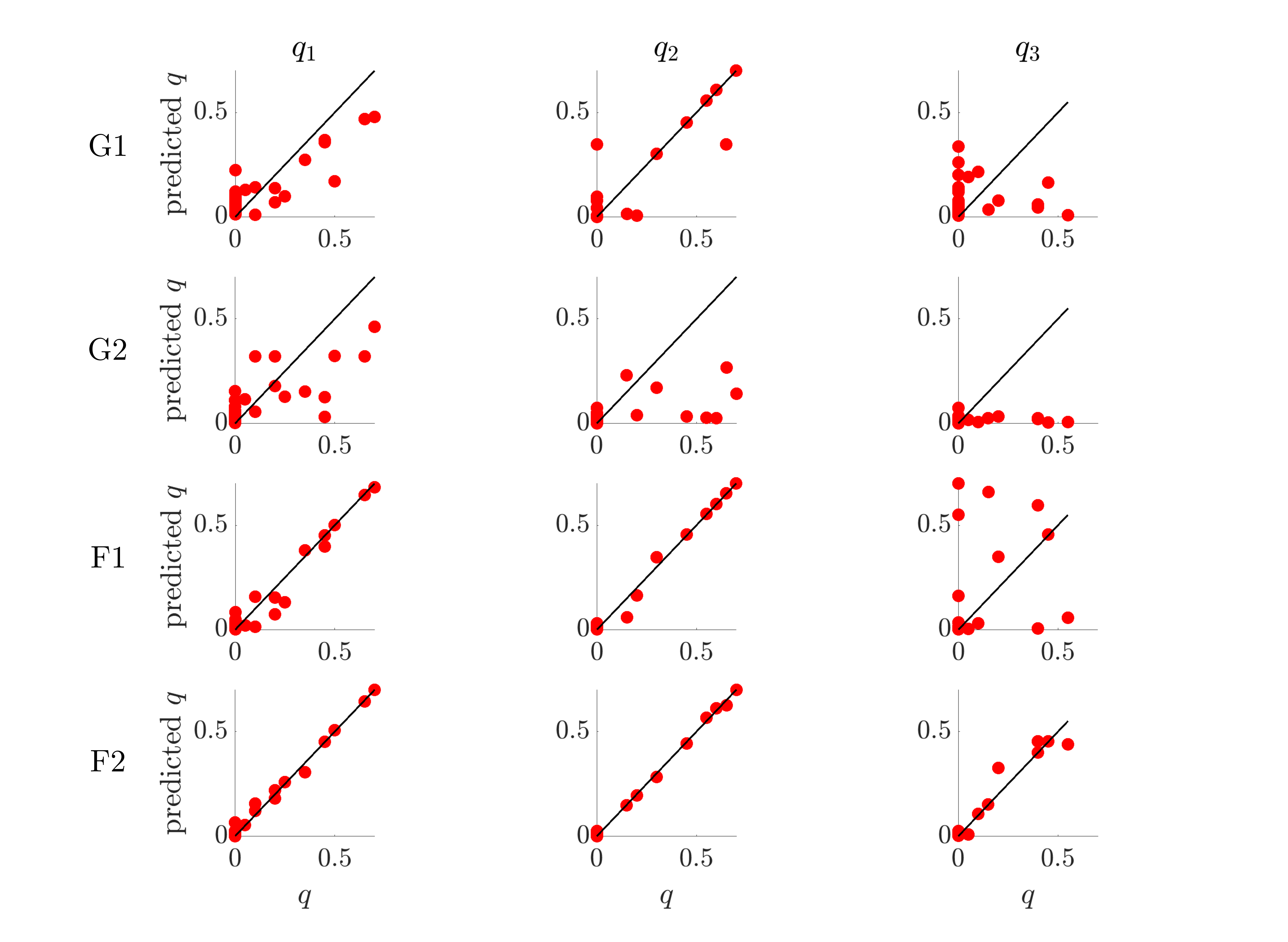}
\caption{Performance of the various features (rows) in one experiment for predicting the three pathological parameters (columns). Ground truth on the horizontal axis, and predicted value on the vertical axis.}
\label{fig:results2}
\end{figure}

Figure \ref{fig:results2} provides a closer look at the results, by focusing on a single experiment where one NN is tested on 1/5 of the dataset (40 noses). 
The specific test set and NN are chosen randomly and are thus representative of the whole set. 
The figure shows the correlation between the predicted (vertical axis) and true (horizontal axis) labels, for each feature and pathology parameter, and emphasizes how the geometric features G1 and G2 do not perform particularly well: it seems that the limited anatomical variability considered in the present work is already enough to throw off the model. It should be noted, though, that the model prediction is not meaningless: non-zero predictions are often associated to non-zero ground truth, in particular for the parameter $q_2$ (hypertrophy of the body of the inferior turbinate), which involves a larger surface. Both G1 and G2 are unable to predict the values of hypertrophy on the middle turbinate. 
Flow features F1 and F2, instead, demonstrate good regression capabilities, especially when used to predict pathologies of the inferior turbinate. For the middle turbinate, the prediction accuracy decreases. Between the two flow features, wall shear stress has a small, but significant edge over pressure.



\section{Conclusions and outlook}
\label{sec:conclusions}

This work has introduced and discussed a novel interaction between Computational Fluid Dynamics (CFD) and Machine Learning (ML). The key conclusion is that CFD-computed information may harbour extremely informative features, and may thus be useful to ML in the execution of classification and regression tasks.

The problem of interest is the flow in the human nose, and the classification of anatomic pathologies, in view of clinical decisions concerning functional surgery of the human upper airways.
In this context, the strategic objective is learning to automatically discriminate physiological inter-subject anatomic variations from variations related to a pathological condition. 
Two major difficulties in this endeavor consist in the large anatomical variability which exists across healthy noses, and in the cost of obtaining the large number of annotated observations that is typically required to train ML algorithms. 
The latter issue, in particular, renders the standard approach of building a deep neural network (which in principle could learn directly from anatomies) highly impractical.

We have shown that an alternate solution strategy is possible: when using features extracted from the flow field computed with CFD, the training of a neural network becomes substantially easier in comparison to equivalent networks that rely on geometry-based features. 
The non-linearity of the Navier--Stokes equations, together with the convective nature of the flow, is such that extracting significant information from the flow field is more effective than looking at the small anatomical changes that are behind that information. 
While a very large number of annotated CT scans (hence, a large amount of purely geometrical information) could in principle lead to a successful ML classification procedure, relying on the computed flow field is an interesting and effective alternative whenever, as in the medical field, annotated CT scans are difficult to obtain, and thus necessarily available in limited quantity.
Albeit the present model only considers localized geometrical deformations, it is found that CFD-based features outperform both small-scale geometrical features like G1 and the large-scale ones conveyed by G2.

The present work and the ensuing conclusions are obviously limited by the extreme simplification of the anatomical model, and by the corresponding low-fidelity CFD approach, based on RANS simulations only. 
It is important to keep in mind that this work does not aim at introducing a clinically usable tool: in a realistic setting, the full parametrization of the entire geometry would be impossible, and alternative ways for representing pathologies would be needed.
However, thanks to the careful design of the reference nose and of its pathologies, which are clinically significant, we are confident that the main conclusions are robust and will continue to apply even when the underlying anatomies become more realistic or, eventually, are derived  from CT scans. 
The results of this work are motivating our ongoing research efforts \citep{schillaci-etal-2022} for the classification of nasal pathologies, where the nose model is substituted with real patient-specific anatomies. Additional difficulties are encountered, like e.g. the need to avoid a full parametrization of the anatomy, but the present results support the design of a procedure based on CFD-computed features.  
At the same time, these conclusions may be of general interest, and pave the way to the use of fluid mechanical features as input to improved ML methods.

\begin{Backmatter}

\paragraph{Acknowledgments}
Computing time was provided by the CINECA Italian Supercomputing Center through the ISCRA-B projects ONOSE-AN and ONOSE-AI.

\paragraph{Declaration of Interests}
The authors declare no conflict of interest.

\paragraph{Ethical Standards}
The research meets all ethical guidelines, including adherence to the legal requirements of the study country.

\paragraph{Data Availability Statement}
Data sets generated during the current study are available from the corresponding author on reasonable request.

\bibliographystyle{jfm}

\end{Backmatter}

\end{document}